\documentclass[10pt,aps,prd,twocolumn,floats,floatfix,showpacs,superscriptaddress,nofootinbib]{revtex4-1}
\usepackage[latin1]{inputenc}
\usepackage{booktabs}
\usepackage{mathtools}
\usepackage{bm}
\usepackage{amsmath,amssymb}
\usepackage[hidelinks]{hyperref}
\usepackage[usenames]{xcolor}
\usepackage{varioref}

\newcommand{\ee}{\mathrm{e}}

\newcommand{\dd}{\mathrm{d}}
\newcommand*\DAlembert{\mathop{}\!\mathbin\Box}
\renewcommand{\rho}{\varrho}
\newcommand{\f}{\varphi}
\renewcommand{\epsilon}{\varepsilon}

\def\beq{\begin{equation}}
\def\eeq{\end{equation}}
\def\bea{\begin{eqnarray}}
\def\eea{\end{eqnarray}}

\def\L{\mathcal{L}}
\DeclarePairedDelimiter{\abs}{\lvert}{\rvert}

\begin{document}
\title{Cosmology with subdominant Horndeski scalar field}

\author{Nicola Franchini}
\affiliation{School of Mathematical Sciences, University of Nottingham,
University Park, Nottingham NG7 2RD, United Kingdom}

\author{Thomas P.~Sotiriou}
\affiliation{School of Mathematical Sciences, University of Nottingham,
University Park, Nottingham NG7 2RD, United Kingdom}
\affiliation{School of Physics and Astronomy, University of Nottingham,
University Park, Nottingham NG7 2RD, United Kingdom}

\begin{abstract}
We study the cosmological evolution of a scalar field in Horndeski gravity, assuming that the scalar field is subdominant with respect to the cosmic fluid. First, we analyse the most general shift-symmetric action that respects local Lorentz symmetry. We show that the bound on the speed of gravitational waves set by GW170817+GRB170817A imposes a constraint only on the linear coupling between the scalar and the Gauss-Bonnet invariant, and this constraint is rather mild. Then, we consider some interesting examples of theories that break shift symmetry, such as the Damour-Esposito-Far\`{e}se model of spontaneous scalarization and a theory with a quadratic coupling to the Gauss-Bonnet invariant that can lead to black hole scalarization. In both cases, tuning of cosmological initial conditions is necessary to keep the scalar field dormant during cosmic evolution.
\end{abstract}

\maketitle

\section{Introduction}
A possible solution to the cosmological constant problem is a modification of general relativity (GR) through the introduction of a new scalar field. Horndeski gravity~\cite{Horndeski:1974wa}, independently rediscovered as generalised Galileon~\cite{Deffayet:2009mn}, has received a lot of attention in this context. It is the most general theory that leads to second order equations upon direct variation. As such, it avoids the Ostrogradski instability that plagues higher-order theories and provides a unified framework for a very large class of scalar-tensor gravity theories.

Cosmology within the Horndeski framework does indeed offer a description of dynamical dark energy models. A late-time accelerated expansion driven by the  scalar field is an attractor for many subclasses of the theory~\cite{Chiba:1999ka,ArmendarizPicon:2000dh,Ferreira:1997hj,DeFelice:2010pv,DeFelice:2010nf,%
Nojiri:2005vv,Koivisto:2006ai,Tsujikawa:2006ph,Nojiri:2005jg,Deffayet:2010qz,Kase:2018iwp,Kase:2018iwp}. However, many of these models predict that the speed of gravitational waves is different from the speed of light~\cite{Lombriser:2015sxa,Lombriser:2016yzn}. The combined detection of gravitational waves and gamma ray bursts originated in the same binary neutron star merger event~\cite{Monitor:2017mdv},  known as GW170817 and GRB 170817A respectively, posed an extremely tight constraint on the speed of gravitational waves
\begin{equation}\label{eq:GWspeed}
-3\cdot10^{-15}\lesssim c_T-1\lesssim7\cdot10^{-16}.
\end{equation}
This bound rules out a very large class of dark energy models within Horndeski gravity~\cite{Ezquiaga:2017ekz,Baker:2017hug,Creminelli:2017sry}. A detailed study of the surviving dark energy models goes beyond the scope of this work and can be found elsewhere~\cite{Ezquiaga:2018btd,Kase:2018aps}.

It should be stressed that the above result is based on the key assumption that the scalar field is the dominant component in the cosmic energy density and drives the accelerated expansion of the Universe. Under this assumption, one can argue that Horndeski theories that satisfy the gravitational wave speed constraints will not be relevant in other regimes, {\em e.g.},  for compact objects~\cite{Tattersall:2018map}. However, if one relaxes the requirement that the scalar is responsible for dark energy, the bound given above does not lead to significant constraints on Horndeski models. The main aim of this paper is to provide a clear demonstration of this last point, which also justifies why certain scalar-tensor theories continue to provide an effective description of new gravitational physics in the strong gravity regime.

Indeed, there is a rich strong gravity phenomenology associated with  theories within the Horndeski class. Solutions with non-trivial scalar configuration exist for neutron stars~\cite{Damour:1993hw,Harada:1997mr,Barausse:2012da,Ramazanoglu:2016kul,Silva:2017uqg} and black holes~\cite{Campbell:1991kz,Mignemi:1992nt,Kanti:1995vq,Yunes:2011we,Sotiriou:2013qea,%
Sotiriou:2014pfa,Doneva:2017bvd,Silva:2017uqg,Antoniou:2017hxj}. Regarding the latter, it is worth noting that no-hair theorems are currently known only for theories that do not exhibit derivative (self-)interactions, {\em e.g.}~\cite{Hawking:1972qk,Bekenstein:1971hc,Sotiriou:2011dz}, or respect shift symmetry~\cite{Hui:2012qt} and do not include a coupling with the Gauss-Bonnet invariant~\cite{Sotiriou:2013qea,Sotiriou:2014pfa}. Moreover, hairy solutions that circumvent the assumptions of no-hair theorems are known even for theories that are covered by them in principle~\cite{Jacobson:1999vr,Torii:2001pg,Cardoso:2013opa,Cardoso:2013fwa,Babichev:2013cya,%
Sotiriou:2013qea,Sotiriou:2014pfa,Herdeiro:2014goa}.

With the motivations laid above, in this work we study the cosmology of a Horndeski scalar field, assuming it is subdominant with respect to the dominant cosmological components (cosmological constant, non-relativistic matter, relativistic matter). The paper is organised as follows: Section~\ref{sec:Horndeski} is devoted to a brief overview of Horndeski gravity and the cosmology derived from it. In Sec.~\ref{sec:ShiftSymmetry} we show the agreement of the shift-symmetric subclass of Horndeski gravity with the standard model of cosmology and we identify a special case: when the scalar field couples to the Gauss-Bonnet scalar. When the shift-symmetry assumption is relaxed, a general study of the cosmology is quite challenging. Thus, in Sec.~\ref{sec:Non-shift} we restrict our attention to some particular models which are also relevant in the strong gravity regime. We conclude with a discussion of the results of this work in Sec.~\ref{sec:Discussion}. Throughout the paper we will assume $4\pi G=c=1$.

\section{Horndeski gravity and cosmology}\label{sec:Horndeski}
One of the leading principles that led Einstein to the equations of GR was the requirement that they are second-order in derivatives. This requirement is necessary to avoid the so-called Ostrogradsky instability. The theorem proven by Ostrogradsky states that non-degerate higher-order theories have Hamiltonians that are not bounded from below. Horndeski gravity~\cite{Horndeski:1974wa,Deffayet:2009mn} is the most general theory which leads to second-order equations of motion for the metric and the scalar field upon direct variation. The action can be expressed in different equivalent representations. Here we report the form of~\cite{Kobayashi:2011nu}
\begin{equation}
\label{eq:horndeski_lagrangian}
	S_H
	=
	\int\dd^4 x\sqrt{-g}(\L_{2} + \L_{3} + \L_{4} + \L_{5})+S_m[\Psi_A,g_{\mu\nu}]
\end{equation}
where $S_m$ is the action for the matter fields $\Psi_A$ and
\begin{align}
\label{eq:horndeski_lagrangian_terms}
	\L_{2} &= G_2(\phi,X)\,, \\
	\L_{3} &= - G_{3}(\phi,X) \Box \phi\,, \\
	\L_{4} &= G_{4}(\phi,X) R + G_{4X}(\phi,X) \left[ \left( \DAlembert \phi \right)^{2} - \left( \nabla_{a} \nabla_{b} \phi \right)^{2} \right]\,, \\
	\L_{5} &= G_{5}(\phi,X) G_{ab} \nabla^{a} \nabla^{b} \phi \\ \notag
           &- \frac{1}{6} G_{5X}(\phi,X) \left[ \left( \DAlembert \phi \right)^{3} - 3 \DAlembert \phi \left( \nabla_{a} \nabla_{b} \phi \right)^{2} + 2 \left( \nabla_a \nabla_b \phi \right)^{3} \right]\,,
\end{align}
where $X = - \frac{1}{2} \nabla_{a} \phi \nabla^{a} \phi$, and $G_{iX}=\partial G_{i}/\partial X$.

For the description of a homogeneous, isotropic and flat Universe we assume Friedmann-Lemaitre-Robertson-Walker metric
\begin{equation}\label{eq:FLRWmetric}
\dd s^2=-\dd t^2+a^2(t)\left[\dd r^2+r^2(\dd\theta^2+\sin^2\!\theta\dd\phi^2)\right],
\end{equation}
where $a(t)$ is the scale factor. The spacetime symmetries require $\phi=\phi(t)$. We assume that the matter action $S_m$ describes a perfect fluid whose energy density is $\epsilon$ and pressure $P$, linked by the equation of state $P=\lambda\epsilon$, with $\lambda$ depending on the fluid considered. A variation of~\eqref{eq:horndeski_lagrangian} with respect to $a(t)$ yields
\begin{equation}\label{eq:friedmannPHI}
6G_4 H^2=\epsilon+\epsilon_\phi,
\end{equation}
where
\begin{align}\label{eq:friedmann1}
  \epsilon_\phi= & 2XG_{2X}-G_{2} + 6X\dot{\phi}HG_{3X}-2XG_{3\phi} \nonumber\\
  + & 24H^2X(G_{4X}+XG_{4X\!X}-12HX\dot{\phi}G_{4\phi X}-6H\dot{\phi}G_{4\phi})  \nonumber\\
  + & 2H^3X\dot{\phi}(5G_{5X}+2XG_{5X\!X})  \nonumber\\
  - & 6H^2X(3G_5\phi+2XG_{5\phi X})
\end{align}
is the field energy density. A variation of~\eqref{eq:horndeski_lagrangian} with respect to $\phi(t)$ gives the equation of motion of the scalar field
\begin{equation}\label{eq:scalarfield}
\frac{1}{a^3}\frac{\dd}{\dd t}\left(a^3J\right)=P_\phi,
\end{equation}
where
\begin{multline}\label{eq:current}
J=\dot{\phi}G_{2X}+6H XG_{3X}-2\dot{\phi}G_{3\phi} \\ +6H^2\dot{\phi}(G_{4X}+2XG_{4X\!X})-12HXG_{4\phi X} \\ +2H^3X(3G_{5X}+2XG_{5X\!X})-6H^2\dot{\phi}(G_{5\phi}+XG_{5\phi X}),
\end{multline}
and
\begin{multline}\label{eq:source}
P_\phi=G_{2\phi}-2X\left(G_{3\phi\phi}+\ddot{\phi}G_{3\phi X}\right) \\ +6\left(2H^2+\dot{H}\right)G_{4\phi}+6H\left(\dot{X}+2HX\right)G_{4\phi X} \\ -6H^2XG_{5\phi\phi}+2H^3X\dot{\phi}G_{5\phi X}.
\end{multline}

For completeness we report how the speed of a gravitation wave is modified in Horndeski cosmology. Indeed, we will apply the constraint~\eqref{eq:GWspeed} to all the models we will take into account. The velocity of gravitational waves is modified as~\cite{DeFelice:2011bh}
\begin{equation}\label{eq:speedGW}
c_T^2=1+\alpha_T,
\end{equation}
where
\begin{equation}\label{eq:alphaT}
\alpha_T=\frac{X\left[2G_{4X}-2G_{5\phi}-(\ddot{\phi}-\dot{\phi}H)G_{5X}\right]}{G_4- 2XG_{4X}+XG_{5\phi}-\dot{\phi}H X G_{5X}}.
\end{equation}

\section{Shift-symmetric theories}\label{sec:ShiftSymmetry}
As a first case, we study the shift-symmetric part of Horndeski action, namely when $G_i(X,\phi)=G_i(X)$. This ensures invariance under shifts of the scalar field $\phi\rightarrow\phi+C$~\cite{Sotiriou:2014pfa}. Equation~\eqref{eq:scalarfield} then becomes
\begin{equation}\label{eq:scalarShiftS}
\frac{\dd}{\dd t}(a^3 J)=0,
\end{equation}
which means $J\propto a^{-3}(t)$.

\subsection{A simple example}
The first model we consider is the simplest one that has a non-trivial $\alpha_T$. It is obtained by choosing $G_2=4X$, $G_4=1+\alpha X$, and $G_3=G_5=0$. With this choice of the Horndeski functions, we have
\begin{equation}\label{eq:J1}
J=2(2+3\alpha H^2)\dot{\phi}.
\end{equation}
If one assumes that 
\begin{equation}\label{eq:alphaAssumption}
\alpha H^2 \ll 1
\end{equation}
then $\dot{\phi} \approx C/a^3(t)$. Note that $\alpha$ has dimensions of a length square, while $H^{-1}$ roughly describes the typical size of the Universe. Hence, Eq.~\eqref{eq:alphaAssumption} can be interpreted as the assumption that the characteristic scale associated in the new coupling of the theory is much smaller than the Hubble scale. It is worth emphasizing that $\alpha$ would be expected to be order ${\rm km}^2$ for a model to have appreciable deviations from general relativity in the strong field.

The energy density of the scalar field reads
\begin{equation}\label{eq:EF1}
\epsilon_\phi = \dot{\phi}^2+\frac{9}{2}\alpha H^2\dot{\phi}^2 \approx \dot{\phi}^2.
\end{equation}
Since  $\dot{\phi}^2$ scales like $a^{-6}$, the scalar field contribution is negligible with respect to every standard Universe content (cosmological constant, dust, relativistic matter). Hence, one gets the standard equation $3H^2\approx\epsilon$ in the late Universe, which is our focus in this paper.

Indeed, it is late Universe cosmology that is relevant for the constraint given in Eq.~\eqref{eq:GWspeed}, as the binary neutron star event happened at $42\,\text{Mpc}$ away from Earth~\cite{Monitor:2017mdv}, well inside the region of a cosmological constant dominated Universe. Hence, using the approximation outlined before, one gets that the deviation of the speeds of GWs from the speed of light is
\begin{equation}\label{eq:aT1}
 \alpha_T\approx \alpha\epsilon_\phi.
\end{equation}
That is, for a given theory (fixed $\alpha$), any constraint on speed can be roughly interpreted as a constraint on the energy density of the scalar during the time interval between emission and detection. Choosing $\epsilon_\phi$ to have a sufficiently small value to evade the GW speed constraint is hence a fully consistent option when $\alpha H^2 \ll 1$, as $\epsilon_\phi$ scales like $a^{-6}$.

One might be worried that our assumption would not be valid any more once one pushes the theory even more backwards in time. However, this would end up in just choosing the initial conditions for the scalar field. Nonetheless, since we are interested in late-time cosmology, this discussion goes beyond the scope of this paper.

See also Ref.~\cite{Diez-Tejedor:2018fue} for another example of cosmology with a subdominant scalar in a subclass of the shift-symmetric Horndeski action.

\subsection{Shift-symmetric theories admitting GR solutions}
We now want to generalise the results of the previous section. To compute the evolution of the scalar field, we need the current
\begin{multline}\label{eq:currentSS}
J=\sqrt{2\abs{X}}G_{2X}+6H XG_{3X} \\ +6H^2\sqrt{2\abs{X}}(G_{4X}+2XG_{4XX}) \\ +2H^3X(3G_{5X}+2XG_{5XX}).
\end{multline}
GR is recovered if $J=0$ when $X=0$. We will assume that a constant scalar solution, for which $X=0$, is admissible. Following Ref.~\cite{Saravani:2019xwx}, we can define the following functions
\begin{align}\label{eq:LLSA}
F_{(2,1)}(X)=\abs{X}^{1/2}G_{2X},& \qquad F_{(3,1)}(X)=XG_{3X},\notag\\
F_{(4,1)}(X)=\abs{X}^{1/2}G_{4X},& \qquad F_{(4,2)}(X)=\abs{X}^{3/2}G_{4X\!X},\notag\\
F_{(5,1)}(X)=XG_{5X},& \qquad F_{(5,2)}(X)=X^2G_{5X\!X}.
\end{align}
The requirement of regularity of $J$ around $X=0$ translates into the requirement of regular expansion of the functions $F_i(X)$. Expanding $F_i(X)$ around $X=0$ implies
\begin{align}
  J= & j_1\dot{\phi}+j_2\dot{\phi}^2+\dots, \label{eq:Jexpansion} \\
  \epsilon_\phi= & \epsilon_2 \dot{\phi}^{2}+\epsilon_3 \dot{\phi}^{3}+\dots, \label{eq:epsiexpa}
\end{align}
with $j_i$, $\epsilon_i$ constants. Note that we have set $G_2(0)\equiv 0$, in order to avoid a shift of the cosmological constant.

The dominant term of the current $J$ is proportional to $\dot{\phi}$, which means $\dot{\phi}=C/a^3(t)$. The latter implies $\ddot{\phi}=-3H\dot{\phi}$. The scalar field energy density~\eqref{eq:epsiexpa} is negligible in the Friedmann equation under our assumptions.

With this behaviour of the scalar field, one gets that the dominant term of $\alpha_T$ decreases as $a^{-6}$, as one can read from the equality
\begin{equation}
\alpha_T=\frac{\alpha_2\dot{\phi}^2+ \alpha_3\dot{\phi}^3+\dots}{1+\beta_2\dot{\phi}^2+\beta_3\dot{\phi}^3+\dots},
\end{equation}
where again $\alpha_i$ and $\beta_i$ are constants.

Note that, as it has been shown in Ref.~\cite{Saravani:2019xwx}, theories that do not admit regular expansion of the functions $F_i(X)$ are either Lorentz-violating, or they can be obtained by adding a linear coupling between $\phi$ and the Gauss-Bonnet invariant $\mathcal{G}=R_{abcd}R^{abcd}-4R_{ab}R^{ab}+R^2$ to the theories that admit a regular expansion in $X$. We will not consider Lorentz-violating theories here and we will discuss the effect of the $\phi \mathcal{G}$ coupling in the next section.

\subsection{Linear coupling to Gauss-Bonnet}\label{sec:linGB}
A linear coupling between $\phi$ and the Gauss-Bonnet invariant is known to be the only term that respects shift symmetry and leads to black hole hair~\cite{Sotiriou:2013qea}. It also features prominently in the classification of theories presented in Ref.~\cite{Saravani:2019xwx}, as the only term that prevents Horndeski theories from admitting all GR solutions without threatening Lorentz symmetry. The simplest action that contains this term is
\begin{equation}\label{eq:GBactionLIN}
S_{\mathcal{G}}=\int\dd^4x\sqrt{-g}\left(\frac{R}{2}+X+\alpha\phi\mathcal{G}\right)\,.
\end{equation}
We emphasize that, since $\mathcal{G}$ is a total derivative, this action is shift-symmetric up to a boundary term. In the notation of Eq.~\eqref{eq:horndeski_lagrangian_terms}, the action \eqref{eq:GBactionLIN} is equivalent to $G_2=X$, $G_4=1/2$, $G_5=-4\alpha\log X$ and $G_3=0$, which is indeed explicitly shift-symmetric. The corresponding current in our cosmological setup is
\begin{equation}\label{eq:scalarfieldGB}
J=\dot{\phi}-8\alpha H^3,
\end{equation}
which gives the solution
\begin{equation}\label{eq:solphiGB}
\dot{\phi}=8\alpha H^3+\frac{C}{a^3(t)}.
\end{equation}
For a matter-dominated Universe, $\dot{\phi}$ has the usual $1/a^3$ behaviour at leading order. However, for a $\Lambda$-dominated Universe there is a substantial difference: the first term can lead to linear growth with time.

Let us analyse this case in mode detail. Substituting the relation~\eqref{eq:solphiGB} into Friedmann equation, one gets at leading order
\begin{equation}\label{eq:EF2}
3H^2\approx\epsilon-\frac{160}{3}\alpha^2 H^6.
\end{equation}
If we assume that $\alpha\ll\frac{1}{H^2}$ (this will be justified later), then Eq.~\eqref{eq:EF2} reduces to the standard Friedmann equation.

In what regards the speed of gravitational waves, differently from the previous cases where the measurement \eqref{eq:GWspeed} poses restrictions only on the initial value of the scalar field, in this case one can obtain a constraint on a free parameter of the theory. Indeed, using eq.~\eqref{eq:solphiGB}, we obtain
\begin{equation}\label{eq:aT2}
\alpha_T\approx-\frac{64\alpha^2H^4}{1+64\alpha^2 H^4}.
\end{equation}
From \eqref{eq:GWspeed} one can infer that
\begin{equation}\label{eq:GBconstraint}
\abs{\alpha H^2}\ll10^{-9},
\end{equation}
which, restoring the units of measure, for the current value of the Hubble constant is equivalent to
\begin{equation}\label{eq:GBconstraint2}
\sqrt{\alpha}\ll 10^{19}\,\text{km}.
\end{equation}
This constraint on the linear coupling between the scalar field and the Gauss-Bonnet invariant is much weaker  than  other known constraints~\cite{Yagi:2012gp,Seymour:2018bce,Maselli:2014fca,Maselli:2017kic,Witek:2018dmd}. Indeed, $\sqrt{\alpha}$ is the length scale associated with the Gauss-Bonnet coupling, whereas $1/H$ is a rough estimate of the size of the Universe. In other words, it appears to be easy to suppress the effect of the linear growth and still have a theory that deviates from GR in the strong curvature regime.

\section{Abandoning shift symmetry}\label{sec:Non-shift}
For shift-symmetric theories, the classification of Ref.~\cite{Saravani:2019xwx}, together with the fact that $\phi$ can be set to zero for constant $\phi$ solutions without loss of generality, helped us to organize our analysis and obtain general conclusions for all theories that respect local Lorentz symmetry. Trying to obtain general conclusions for theories that do not respect shift symmetry is more challenging. Instead, we will focus here on some specific examples of particular interest for the strong field regime.

\subsection{Mass term}
The simplest term that breaks shift symmetry is a mass term, which we include here for completeness. We have $G_2=X-m^2\phi^2/2$, $G_3=0$, $G_4=1/2$, $G_5=0$. The equation of motion for the scalar field is
\begin{equation}\label{eq:massivephi}
\ddot{\phi}+3H\dot{\phi}+m^2\phi=0,
\end{equation}
whose general solution is
\begin{equation}\label{eq:solmassive}
\phi(t)=\ee^{-\frac{3}{2}H t}\left(C_1\ee^{-\omega_0 t}+C_2\ee^{\omega_0 t}\right),
\end{equation}
where
\begin{equation}\label{eq:omgamass}
\omega_0=\frac{1}{2}\sqrt{9H^2-4m^2}.
\end{equation}
The dominant term in the scalar field energy density goes as
\begin{equation}\label{eq:massivedensity}
\epsilon_\phi\approx\tilde{C}_2\ee^{-t\left(3H-2\omega_0\right)},
\end{equation}
which is also decreasing with time, since $3H-2\omega_0$ is always greater than $0$ for real $m$. GR is an attractor.

\subsection{Damour-Esposito-Far\`{e}se gravity}
Damour-Esposito-Far\`{e}se (DEF) gravity~\cite{Damour:1992kf,Damour:1993hw} is a particular class of scalar-tensor gravity (which is a generalizaton of Brans-Dicke theory of gravitation \cite{Brans:1961sx}), with $G_2=X\frac{\omega(\phi)}{\phi}$, $G_4=\phi/2$ and $G_3=G_5=0$.  This choice of the Horndeski functions sets the theory in the so-called Jordan frame. One can perform a conformal transformation of the metric, together with a field redefinition
\begin{equation}\label{eq:conformaltransformation}
g_{\mu\nu}=A^2(\f)g^*_{\mu\nu}, \qquad \phi=A^{-2}(\f),
\end{equation}
to bring the action in the Einstein frame, where $G_2=2X$, $G_4=1/2$, $G_3=G_5=0$ and $S_m=S_m[\Psi_A,A^2(\f)g^*_{\mu\nu}]$, with $\Psi_A$ the generic matter fields. This transformation requires also the further identification $2\omega(\phi)=1/\alpha^2(\f)-3$, where $\alpha(\f)=A'(\f)/A(\f)$. In this reference frame, the Friedmann equation is
\begin{equation}\label{eq:friedmannDEF}
3H_*^2=\epsilon+\dot{\f}^2,
\end{equation}
where $H_*=\dot{a}_*/a_*$ is the Hubble parameter in the Einstein frame, while the scalar field equation reads
\begin{equation}\label{eq:scalarfieldDEF}
\ddot{\f}+3H_*\dot{\f}=-\frac{\alpha(\f)}{2}\epsilon(1-3\lambda).
\end{equation}
With a time coordinate redefinition $\dd\tau\equiv H_*\dd t_*$, Eq.~\eqref{eq:scalarfieldDEF} becomes
\begin{equation}\label{eq:scalarfieldDEF2}
\frac{2}{3-\f'^{2}}\f''+(1-\lambda)\f'=-(1-3\lambda)\alpha(\f),
\end{equation}
where a prime is a derivative with respect to $\tau$. This equation describes the oscillations of a field with velocity dependent mass $2/(3-\f'^{2})$, a friction term $(1-\lambda)\f'$ and an external force which depends on the form of $\alpha(\f)$.

The evolution of this equation for $\alpha(\f)=\beta\f$ has been studied in~\cite{Damour:1992kf,Damour:1993id}. When $\beta>0$, the solution of Eq.~\eqref{eq:scalarfieldDEF2} in a matter-dominated Universe for $\f$ is a decaying exponential. On the other hand, the $\beta<0$ case provides a runaway solution which is in contrast with Solar System constraints. In~\cite{Anderson:2016aoi} the authors provide a similar analysis generalising the function $\alpha$ as a polynomial in $\f$. They show that a large class of these theories have solutions which agree with Solar System constraints.

The cosmological evolution of the scalar field in DEF theory is particularly relevant for neutron star physics. Indeed, it is well known that this theory leads to spontaneous scalarization of neutron stars: a phase transition which occurs to the scalar field when the compactness of the star exceeds a certain threshold. For example, in the $\alpha=\beta\f$ case, spontaneous scalarization takes place for ultra-dense stars when $\beta\lesssim-4.5$.

This model of spontaneous scalarization shows a contrast with cosmological evolution: for the choice of parameters that do not allow spontaneous scalarization, one gets a cosmological growth of the scalar field consistent with Solar System constraints. On the other hand, one can force the scalar to respect the constraints only with a fine-tuning of the initial data. See Ref.~\cite{Anderson:2016aoi} for a detailed discussion of the required tuning when $\alpha$ has a polynomial scaling. A possible way out is to consider a massive scalar~\cite{Alby:2017dzl}.

\subsection{Scalar-Gauss-Bonnet}
A theory which received a lot of interest recently is the scalar-Gauss-Bonnet (sGB) class. This is a generalisation of the shift-symmetric theory studied in Sec.~\ref{sec:linGB}. The linear coupling is here substituted with a function $f(\phi)$ \cite{Kanti:1995vq,Antoniou:2017acq,Doneva:2017bvd,Silva:2017uqg}:
\begin{equation}\label{eq:GBaction}
S_{GB}=\int\dd^4x\sqrt{-g}\left(\frac{R}{2}+X+f(\phi)\mathcal{G}\right).
\end{equation}
This theory is a subclass of Horndeski gravity, and one can recast it into the action~\eqref{eq:horndeski_lagrangian} through the following transformation~\cite{Kobayashi:2011nu}:
\begin{align}\label{eq:GRtransformation}
  G_2= & X+8f_{4\phi}\,X^2(3-\log X), \\
  G_3= & 4f_{3\phi}\,X(7-3\log X), \\
  G_4= &\frac{1}{2}+4f_{2\phi}X(2-\log X), \\
  G_5= & -4f_{\phi}\log X,
\end{align}
where $f_{n\phi}=\partial^n f/\partial\phi^n$.

The equation of motion of the scalar field reads
\begin{equation}\label{eq:scalarfieldsGB}
\DAlembert\phi=-f_{\phi}\mathcal{G}.
\end{equation}
We stick to the requirement that the theory admits solutions of GR. This requires that a constant value for the scalar $\phi_0$ exists, such as
\begin{equation}\label{eq:GRcondition}
f_\phi(\phi_0)=0.
\end{equation}
From the perspective of compact objects, it was shown that, if
\begin{equation}\label{eq:haircondition}
f_{2\phi}\mathcal{G}>0,
\end{equation}
then black holes (or neutron stars) can spontaneously grow scalar hair~\cite{Doneva:2017bvd,Silva:2017uqg}. This mechanism is analogous to the spontaneous scalarization of neutron stars presented before.

Motivated by these results, in this section we focus on the cosmological viability of this class of theories. For illustrative purposes, we chose the simplest coupling function, a quadratic coupling $f(\phi)=\beta\phi^2/8$~\cite{Silva:2017uqg}. In terms of the Horndeski functions, this becomes $G_2=X$, $G_3=0$, $G_4=1/2+\beta(2-\log X)$ and $G_5=-\beta\phi\log X$. One would need to supplement our choice of $f$ with higher order $\phi$ corrections for scalarized black hole solutions to be stable~\cite{Blazquez-Salcedo:2018jnn,Minamitsuji:2018xde,Silva:2018qhn}, but  $f(\phi)=\beta\phi^2/8$ is the simplest choice that captures the onset of the instability that leads to scalarization~\cite{Silva:2017uqg}. The condition \eqref{eq:haircondition} translates into $\beta>0$.

The equation of motion for the scalar field on a $\Lambda$-dominated cosmological background is
\begin{equation}\label{eq:scalarfield^2GB}
\ddot{\phi}+3H\dot{\phi}-6H^4\beta\phi=0,
\end{equation}
whose solution is
\begin{equation}\label{eq:solphi^2GB}
\phi=\ee^{-\frac{3}{2}Ht}\left(C_1\ee^{-\omega_0 t}+C_2\ee^{\omega_0 t}\right),
\end{equation}
where $C_{1,2}$ are integrating constants and
\begin{equation}\label{eq:omegaGB}
\omega_0=\frac{H}{2}\sqrt{9+8\beta H^2}.
\end{equation}
In the case where $\abs{\beta H^2}\ll1$, we can approximate as
\begin{equation}\label{eq:omegaGB2}
\omega_0\approx\frac{3H}{2}\left(1+\frac{4\beta H^2}{9}\right).
\end{equation}
This approximation is consistent with the assumption that any modification to gravity given by the scalar field happens at length scales much smaller than those of cosmology (see the discussion at the end of Sec.~\ref{sec:linGB}).
At leading order, the Friedmann equation is modified by the factor
\begin{equation}\label{eq:GBphi^2friedmann}
\epsilon_\phi\approx\frac{142}{9}C_2^2\beta^2H^6\ee^{\frac{4}{3}\beta H^3t}.
\end{equation}
Here we have two cases: if $\beta<0$, then the exponent is negative, and thus the late-time expansion is not affected by the presence of the scalar field; if $\beta>0$, then the scalar field energy density is increasing exponentially with time. However, we notice that in the limit $\abs{\beta H^2}\ll1$, the contribution of $\epsilon_\phi$ to the Friedmann equation is negligible, unless $\abs{C_2}\gg1/\abs{\beta H^2}$.

The comparison with the DEF case is immediate. The choice of $\beta>0$, which from Eq.~\eqref{eq:haircondition} is necessary to have spontaneous scalarization in quadratic sGB gravity, leads to a cosmology that requires tuning of initial data to be in agreement with experimental constraints. For example, the modification factor to the speed of GWs is
\begin{equation}\label{eq:aTGB2}
\alpha_T=\frac{16\beta^2 H^4(4\beta H^2-3)C_2^2}{9\ee^{-\frac{4}{3}\beta H^3 t}+48\beta^2H^4C_2^2}.
\end{equation}
Assuming $\abs{\beta H^2}\ll1$, and imposing \eqref{eq:GWspeed} one gets
\begin{equation}\label{eq:GB2constraint}
\sqrt{\beta C_2}\ll10^{19}\,\text{km}.
\end{equation}
Hence, one can satisfy the gravitational wave speed bound by choosing $C_2$ to be sufficiently small. A key difference from the DEF case is that the tuning required is milder here. One way to see this is through Eq.~\eqref{eq:GBphi^2friedmann}, where $\beta H^2$ appears both as a multiplicative factor and in the exponent.

Tuning of initial data might be undesirable in general, but it is in principle possible when it comes to late time evolution. In the inflationary period instead, at least in the simplest scenario, a quasi-de Sitter phase with a large effective cosmological constant would enhance very significantly the level of tuning needed to avoid the tachyonic instability. It has recently been argued in~\cite{Anson:2019uto} that quantum fluctuations would suffice to source the instability. Without a concrete model of inflation, it is hard to make more robust statements. The required tuning even at the purely classical level is enough to suggest that both the DEF and the Gauss-Bonnet scalarization models need to be embedded within the framework of a larger theory if they are to be valid effective field theories all the way from the early universe to compact objects.

\section{Discussion}\label{sec:Discussion}
In this work we have analysed the cosmology of a scalar field in Horndeski gravity, with the assumption that it is subdominant with respect to the standard $\Lambda$CDM evolution of the Universe. At first, we focused on shift-symmetric theories. We checked that this assumption is not in contrast with any of the theories which admit vacuum GR solutions. Indeed, every theory in this subclass is consistent with a cosmological constant-dominated Universe, with a non-zero, yet very small, energy density of the scalar degree of freedom. In fact, one can also easily verify that the corresponding solutions satisfy the conditions of Ref.~\cite{DeFelice:2011bh} for stability under scalar and tensor perturbations. 

Then, we moved to linear Gauss-Bonnet gravity. The linear coupling between the scalar and the Gauss-Bonnet invariant is the only shift-symmetric term that is consistent with local Lorentz symmetry and yet prevents the shift-symmetric theory from admitting all of the solutions of GR~\cite{Saravani:2019xwx}. It is also the only term that leads to black hole hair~\cite{Sotiriou:2013qea}, under the assumption of flat asymptotics. Interestingly, it also forces the scalar field to grow ``cosmological hair"  in presence of a cosmological constant,  which in principle  back-reacts in the Friedmann equation, modifying the cosmological evolution. However, this effect can be suppressed by a suitable choice of the coupling constant and the constraint one obtains from gravitational waves propagation is rather weak.

For the non-shift-symmetric theories, we did not perform a systematic analysis of every possible theory, but we focused on a few interesting examples. For example, we re-derived the known result that the DEF model of spontaneous scalarization requires a tuning of the initial data in the cosmological evolution, otherwise scalarization will eventually affect any astrophysical body due to non-trival asymptotics for the scalar field.
We performed a similar analysis for the other known case of spontaneous scalarization within the Horndeski framework, \textit{i.e.} when the scalar field couples to the Gauss-Bonnet invariant. We found that the choice of the parameters which gives scalarization is in principle producing a non-trivial contribution of the scalar field in the energy density of the Universe and tuning is required, as in the case of the DEF model. However, the degree of tuning is lower due to the form of the coupling and it is easier to suppress the effect of the scalar on cosmic evolution. Clearly, any tuning is undesirable and is unlikely to be possible when quantum fluctuation are taken into account~\cite{Anson:2019uto}. This suggests that scalarization models need a suitable completion if they have to make sense as quantum, or even semiclassical, theories.

\section{Acknowledgements}
We would like to thank Mehdi Saravani for fruitful discussions. T.~P.~S.~acknowledges partial support from the STFC Consolidated Grant No.~ST/P000703/1. We would also like to acknowledge networking support by the COST Action GWverse CA16104.

\bibliography{bibnote}

\end{document}